# Analysis of a metallic nano-rod polarizer using finite-difference-time-domain method

Baek-Woon Lee[1], Young-Gu Ju[2*]

[1]*LCD Business Unit ,Samsung Electronics Corporation, San 24, Nongseo-ri, Giheung-up, Yongin-si, Gyeonggi-do, 449-901, Korea*

[2]*Department of Physics Education, KyungPook National University, 702-701, Daegu, Korea*

The polarization behavior of metallic nano-rods is analyzed by finite-difference-time-domain method. When the average spacing between the nano-rods is less than a half wavelength, the layer reflects the light polarized parallel to the nano-rods as in a nano-slit. However, when the spacing is larger than a half wavelength, the metallic surface absorbs the light polarized perpendicular to the rods resulting in a polarization reversal. Multiple layers of nano-rods can make a polarizer with high extinction ratio and good transmittance.

## 1. INTRODUCTION

Polarizer is one of the most frequently used optical components in optics industry. Especially, as liquid crystal display(LCD) becomes more popular in the commercial market like monitor and television, the interest on the high performance polarizer grows rapidly to improve the efficiency of the related devices.

In general, polarizer is made from polyvinyl alcohol plastic with an iodine doping.[1] However, the recent advance of nano-technology opens a way to fabricate various forms of nano structures such as nano-wire and nano-rods. The engineering of the structre in nano-scale results in a manipulation of optical property not by means of material, but by means of strucutural parameters. The nano-structure like a nano-wire grid polarizer is already applied to obtain high performance polarizer for enhancing the brightness of LCD display.[2]

The optical properties of nano-rods are still under investigation since many types of nano-rods were synthesized in the laboratory.[3, 4] Although the theoretical efforts based on effective medium theory are tried and demonstrated partially successful, the clear and simple understanding on the polarization behavior of nano-rods is not available, especially to experimentalists.[5]

In this paper, the authors analyze the polarization behavior of a metallic nano-rod film using finite-difference-time-domain(FDTD) method. The simulation results by changing the parameters including diameter, period, duty cyle and the number of layers would give more intuitive picture of polarization characteristics of nano-rods than the method based on the sophiscated calculus. In addition, the analysis reveals that there exist polarization reversal depending on the structural parameter of nano-rods layers, which is caused by completely different physical mechanisms. Finally, the authors compares a metallic nano-rod polarizer and other types of polarizers in terms of physical origin and suggest the structural range in order to use a metallic nano-rod polarizer with high extinction ratio.

## 2. SIMULATION DETAILS

As mentioned earlier, the authors choose the FDTD in order to study the polarization characteristics of metallic nano-rods. The FDTD is being widely used these days to solve the Maxwell equation in the field of wireless communication and nano-photonics[6, 7]. In principle, FDTD is accurate even in the scale limit smaller than a wavelength, as far as numerical errors and quantum effects are ignorable. Since the nano-rods treated in this paper is normally larger than a tenth of a wavelength, the FDTD approach is presumed to give good numerical data which cannot be obtained by applying the scalar diffraction theory. In the simulation, we used a three-dimensional (3D) FDTD with periodic boundary conditions in the y direction, which saves a great deal of computation time and enables one to scan a wide parameter range compared to nominal 3D calculations. The periodic 3D FDTD assumes that a physical situation has symmetry along one axis in terms of optical structure and dipole sources. For instance, if one simulates a circular profile in this quasi-2D model, it corresponds to an infinitely long cylinder in the y direction. Therefore, the periodic FDTD modeling is an efficient tool for analyzing the polarization behavior of nano-rods with accuracy and speed.

An example of permittivity profile is shown in Fig. 1. The nano-rod film in the example is composed of 10 layers of aluminum nano-rods whose diameter is about 50 nm. The diameter, the period and the number of layers varies depending on the parameter to scan in each calculation. The randomness is given to the position and the diameter of nano-rods to avoid the possible abnormal trait due to the periodic structure because the average coordinates of the nano-rods satisfy a hexagonal lattice. Polarization behavior is basically characterized by the transmission efficiency of a nano-rod film when the plane wave is incident on the layer in normal direction. The transmission is defined as the ratio of the transmitted energy to the incident energy. The energy of a plane wave is measured by integrating Poynting vector along the detector line which is placed in parallel to the layer before and after the wave goes through the block of layers. In the same manner, the reflection and absorption are defined the ratio of the reflected energy and the absorbed energy to the incident energy, respectively. As for the polarization, $E_x$ and $E_y$ polarization exist since the plane wave propagates in z-direction.

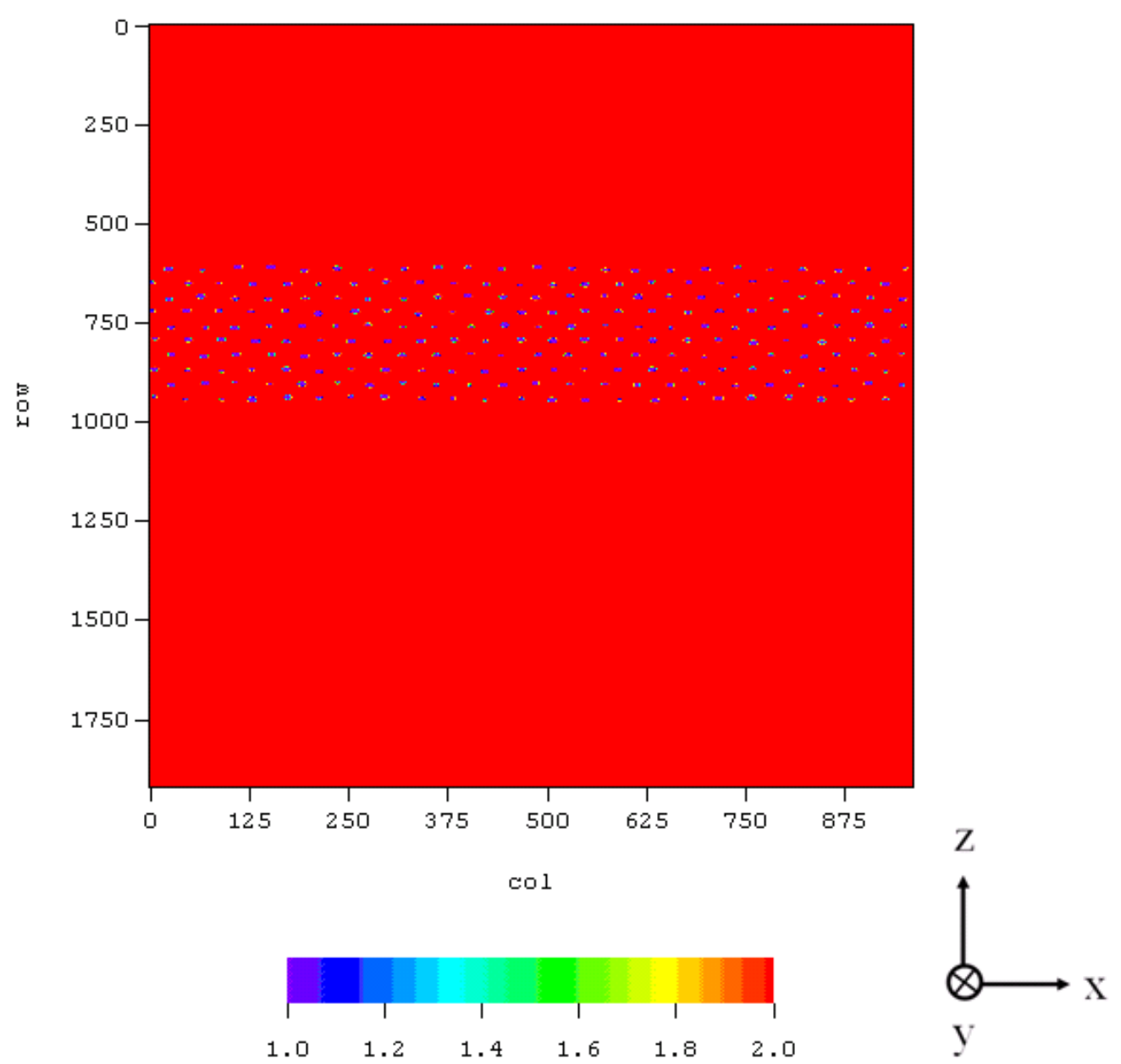

Fig. 1. The cross-section of permittivity profile used in the calculation is displayed. Aluminum nano-rods are placed near hexagonal lattice with randomness in diameter and position. The average diameter of the nano-rods is 50 nm.

To ensure the accuracy of calculation, the meshing size of FDTD is gradually decreased until the transmission converges to a certain value. It is important to use sufficient small mesh in case of the permittivity profile containing metals, otherwise, the simulation may not provide good interpretation even in qualitative manner.

### 3. RESULTS AND DISCUSSION

Polarization analysis of nano-rods starts from a single layer of nano-rods. The period of nano-rods is selected as the first parameter to look into for studying the polarization behavior of a nano-rod layer. When the average spacing between nano-rods are changed around a half wavelength as shown in Fig. 2-(a), the transmission curve shows a polarization reversal around a half wavelength.

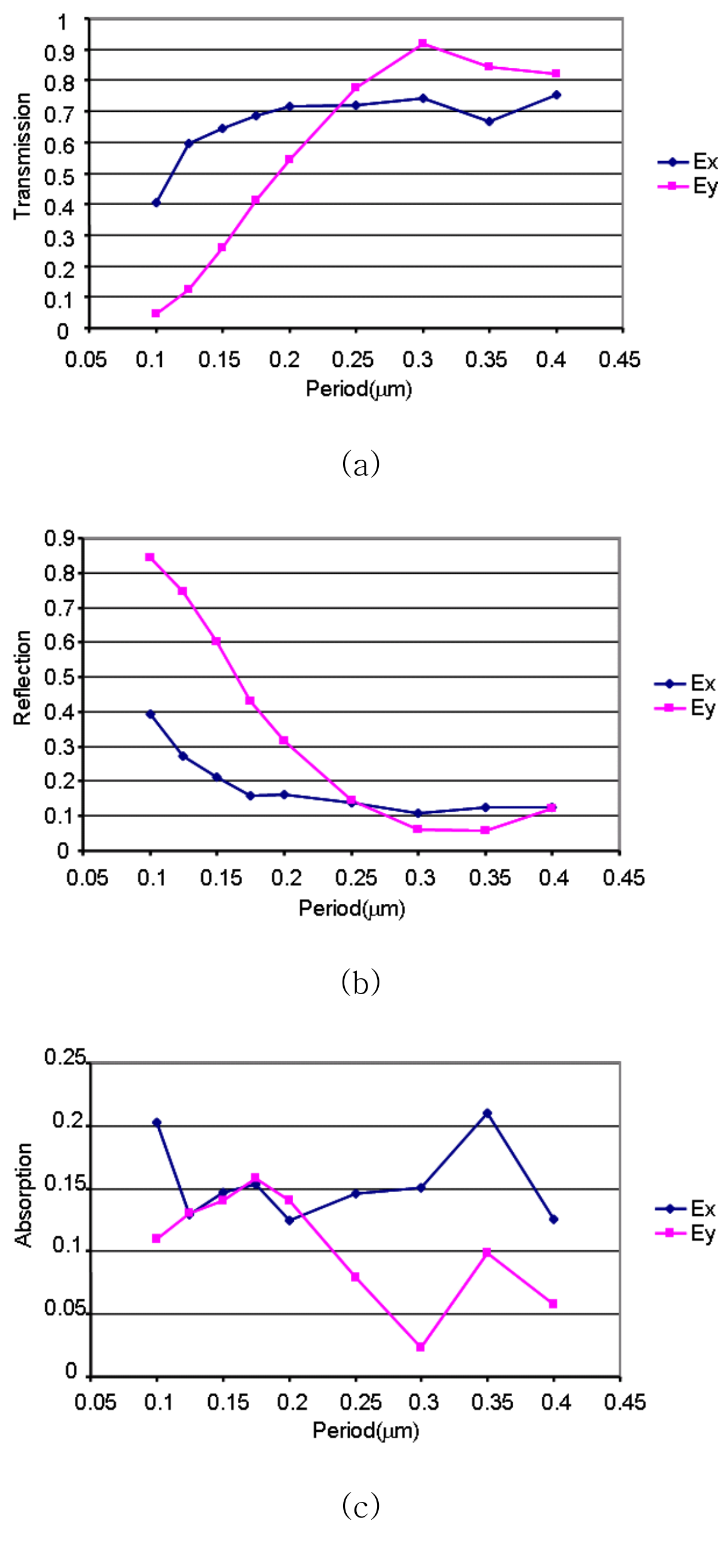

Fig. 2. The transmission (a), reflection (b) and absorption (c) of nano-rod film as a function of period for two polarizations. The wavelength is 460 nm and the average diameter of nano-rods is 50 nm.

Whereas $E_y$ polarization has greater transmission above a half wavelength, the $E_x$ polarization becomes dominant below a half wavelength. The low transmission of $E_y$ polarization can be attributed to the boundary condition at the interface between metal and air. The same phenomenon is observed in a single nano-slit.[6] Since the electric field parallel to the surface is continuous across the boundary and the field inside metal is zero, the $E_y$ should go to zero near metallic surface. On the other hand, the electric field normal to the surface satisfies different boundary condition and doesn't have to go to zero at the surface. Therefore, when the slit opening is narrower than a half wavelength, the optical mode doesn't exist for $E_y$ polarization. The energy blocked by narrow slit opening is reflected as seen in Fig. 2-(b). The reflection efficiency of $E_y$ polarization rises faster than that of $E_x$ polarization as the period decreases.

The polarization behavior in the upper region seems to be governed by different physical mechanism from that of lower region. The dominance of $E_y$ polarization may be caused rather by the absorption process of metallic surface than by the reflection as occurred in the below region. The fractional absorption of $E_x$ polarization is greater than that of $E_y$ polarization when the period is larger than a half wavelength. The difference of reflection efficiencies between the two polarizations in the larger periods is less than the absorption difference as shown in Fig. 2-(b). Thus, the polarization reversal observed at larger period is more likely to be driven by the absorption due to the metallic surface.

A single nano-rod layer exhibits polarization discrimination and polarization reversal around a half wavelength. As the number of layers increase, the polarization distinction ratio increases as seen in Fig. 3. As for the double layer, the $E_x$ polarization has higher transmission efficiency than $E_y$ polarization. The transmission of both $E_x$ and $E_y$ polarizations decreases as the period gets smaller. At the period of 0.15 μm, the transmission of $E_y$ drops nearly to zero and the extinction ratio reaches a high value. On the other hand, at the period of 0.35 μm, the $E_y$ polarization is dominant and the extinction ratio increases with the number of layers. The transmission of the larger period stays at a high value with the increased number of layers, which is not the case for the transmission of the smaller period. For instance, at the period of 0.35 μm, 10 layers of nano-rods have an extinction ratio of 21 and transmission of main polarization is about 60 %, which is twice greater than that of double layer at the period of 0.125 μm.

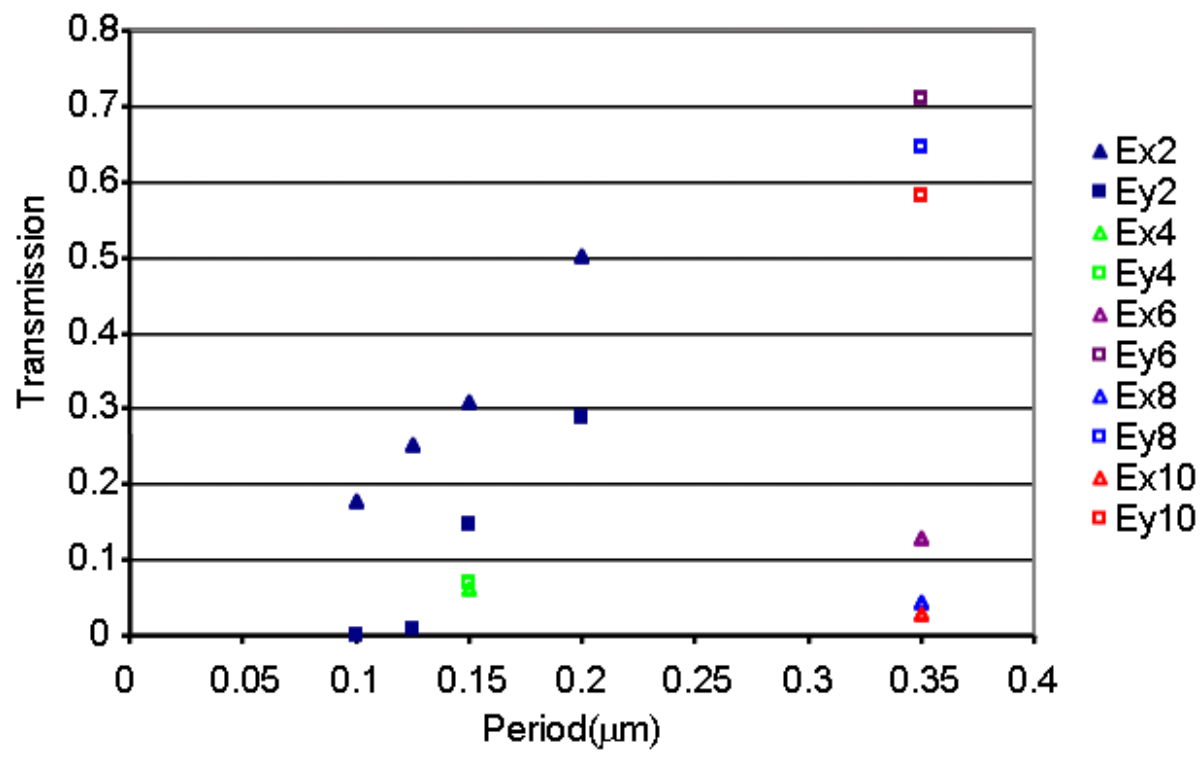

Fig. 3. The transmission of the multiple nano-rod layers is plotted. The average diameter of nano-rods is 50 nm. The suffix indicates the number of layers.

In fact, the traditional polarizer like H-sheet also works as an absorptive mode. However, the physical mechanism seems different because the H-sheet transmits the polarization perpendicular to the polymer chain with an iodine doping. It is known that the electrons from the iodine dopant are able to move along the chains, ensuring that the ligth polarized parallel to the chains is absorbed by the sheet.[1] On the contrary, the nano-rods absorbs the light polarized perpendicular to long axis of the rods. In addition, the polarization discrimination found in the lower period region is similar to that of a wire-grid polarizer since both of them originate from the reflection of a light from a nano-slit with the opening less than a half wave-length.[6]

## 4. CONCLUSION

The polarization behavior of metallic nano-rods is analyzed by FDTD. When the average spacing between the nano-rods is less than a half wavelength, the layer reflects the light polarized parallel to the nano-rods as in a nano-slit. However, when the spacing is larger than a half wavelength, the metallic surface absorbs the light polarized perpendicular to the rods resulting in a polarization reversal. The polarization extinction ratio in a single layer is intensified as the number of the nano-rod layer increases. 10 layers of nano-rods attain the polarization extinction ratio of 21 and transmittance of 60 % if the average diameter is 50 nm and the average period is 0.35 μm at the wavelength of 0.46 μm. The physical mechanism causing the polarization discrimination of nano-rods seems different from those of traditional polarizer like H-sheet.

**Acknowledgment:** This work was supported by the Korea Research Foundation Grant funded by the Korean Government (MOEHRD, Basic Research Promotion Fund) (KRF-2007-331-D00318). This work was supported by Samsung Electronics Corporation.